\documentclass[journal]{IEEEtran}
\ifCLASSINFOpdf

\else

\fi
\usepackage{multirow}

\usepackage{amsmath}
\usepackage{mathtools}
\usepackage{breqn}

\usepackage{algorithm}
\usepackage{algorithmic}
\usepackage{savesym}
\usepackage{graphicx}
\newcommand{\myparagraph}[1]{\smallskip \noindent{\bf {#1}.}}

\begin{document}

\title{Spatial-Temporal Anomaly Detection for Sensor Attacks in Autonomous Vehicles}

\author{Martin~Higgins, 
        Devki Nandan Jha, 
        and~David~Wallom \\
        \textit{Oxford e-Research Center,  University of Oxford,} Oxford, United Kingdom \\
        Martin.Higgins@engs.ox.ac.uk, Devki.Jha@eng.ox.ac.uk, David.Wallom@oerc.ox.ac.uk

}
\maketitle

\begin{abstract}
Time-of-flight (ToF) distance measurement devices such as ultrasonics, LiDAR and radar are widely used in autonomous vehicles for environmental perception, navigation and assisted braking control. Despite their relative importance in making safer driving decisions, these devices are vulnerable to multiple attack types including spoofing, triggering and false data injection. When these attacks are successful they can compromise the security of autonomous vehicles leading to severe consequences for the driver, nearby vehicles and pedestrians. To handle these attacks and protect the measurement devices, we propose a spatial-temporal anomaly detection model \textit{STAnDS} which incorporates a residual error spatial detector, with a time-based expected change detection. This approach is evaluated using a simulated quantitative environment and the results show that \textit{STAnDS} is effective at detecting multiple attack types.
\end{abstract}

\begin{IEEEkeywords}
Autonomous vehicles, cyber-physical, security, LiDAR, ultrasonics
\end{IEEEkeywords}
\IEEEpeerreviewmaketitle

\section{Introduction}
\IEEEPARstart{A}{utonomous} vehicles (AVs) offer many opportunities for efficiency, convenience and cost savings to humanity going forward. In order to allow for self-driven transit, AVs rely on numerous sensing subsystems including cameras, GPS and ToF sensors (e.g., LiDAR, ultrasonic sensors and radar). These sensors collect various data so that the perception algorithm executed by the AVs can bestow awareness of the surrounding environment. This awareness is important when making decisions for safety-critical tasks such as detecting traffic signals and avoiding pedestrians/obstacles. 
The ToF sensors act as an important tool in both self-driving as well as conventional vehicle navigation by allowing the vehicle to reliably ascertain distance. 

Research also shows that AVs by their nature are highly prone to cyber attacks as compared to assisted- or driver-piloted vehicles \cite{Thing2017AutonomousDefences,Sun2022ACAVs,Kim2021CybersecurityDefense}. With the increased reliance on the sensing subsystem, AVs are susceptible to numerous cyber attacks.   Due to the safety-critical nature of AVs, any ToF data error or misinterpretation can lead to fatal consequences \cite{Kohli2018EnablingCrash,Hong2021AICars}. There are numerous attack surfaces for AV perception, an attacker can alter the collected ToF sensor data to either introduce a fake object or hide the actual object, either of which can lead to catastrophic consequences.

Despite the ubiquity of ToF sensing mechanisms in modern vehicles, only a few works have examined the potential attacks they could possess \cite{PetitRemoteLiDAR,Pham2021AVehicles,Li2021FoolingPerturbation,Cao2019AdversarialDriving,Kamal2021GPSVehicles,Hallyburton2021SecurityVehicles}. However, the attack type defined in these works are specific to a single type and is not always prevalent. There also exists some works which consider the attack detection mechanisms \cite{Alheeti2022LiDARVehicles,Sun2020TowardsCountermeasures,Wang2022DetectionValidation}. These mechanisms offer specialised use cases and cannot be generalised for any type of ToF attack defence. 

In this work, we examine some simple but potentially fatal spoofing style attacks against ToF sensing devices and propose a spatial-temporal anomaly detection scheme (STAnDS) to defend against these kind of attacks. 

\subsection{Novel Contributions}
This work outlines a spatial-temporal anomaly detection scheme (STAnDS) for attacks against ToF sensors. The contributions for this work are as follows:
\begin{itemize}
        \item To start, we formulate the spatial integrity test for sensor measurement corroboration. The residual error-based approach uses sensors measuring the same object distance from different locations and the standard measurement error to verify measurement integrity. 
        \item The spatial test is then supplemented by a temporal, expected change-based detection approach which allows the system operator to evaluate unlikely system changes which may be due to attack intervention. 
        \item We simulate this approach under attack vectors and normal operating conditions. We explore specifically the impact of triggering attacks and deflection-style attacks on our detection approach.     
\end{itemize}

 \myparagraph{Outline} The rest of the paper is structured as follows. Section \ref{sec:related} discuss recent related work and presents the motivation for the proposed work. Section \ref{sec:formal} discusses the formulation of attack detection approach and  the basic implementation details of our proposed framework, \textit{STAnDS}. Evaluation of \textit{STAnDS} is presented in Section \ref{sec:evaluation}.  Finally the paper is concluded with some future work in Section \ref{sec:conclusion}.


\section{Related Work}
\label{sec:related}
LiDAR is one of the most commonly utilised ToF sensors in AVs. Several previous works have shown how LiDAR can be vulnerable to spoofing attacks. In \cite{PetitRemoteLiDAR}, remote attacks on LiDAR and camera-based sensing are explored wherein the authors show it is possible to confuse the field of vision of LiDAR using lasers to spoof the LiDAR sensors. In this work, Petit \textit{et al.} show that it is possible to hide objects from the LiDAR sensor using these types of attacks. In \cite{Pham2021AVehicles} Pham \textit{et al.} implements a spoofing style attack against LiDAR systems. This attack is shown to be able to simulate the existence of an object in the field of view. The secondary layer vulnerabilities of LiDAR are discussed in \cite{Li2021FoolingPerturbation} where the LiDAR motion compensation is exploited to create an adversarial trajectory vector which in turn could cause collisions or otherwise alter the navigation of the AVs. In \cite{Cao2019AdversarialDriving}, adversarial sensor attacks against LiDAR are examined and demonstrated in two case studies, one instigated braking and another which forces the car to remain stationary both using LiDAR spoofing. 

To address some of the previously discussed attacks, some works have attempted to create models capable of detecting LiDAR attacks. In \cite{Alheeti2022LiDARVehicles}, a decision tree-based approach for attack detection is analyzed. In \cite{Shafique2021DetectingModels} a machine learning based approach to model detection is taken and applied in UAV style systems. In \cite{Sun2020TowardsCountermeasures} a series of countermeasures to enhance LiDAR perception are explored. They construct Black Box style spoofing attacks as well as suggesting a general architecture for embedding the front view of the LiDAR system. In \cite{Cao2021Demo:Targets} a real-life LiDAR spoofing attack is simulated on a moving target to analyse how effective LiDAR spoofing might be in practice. In \cite{Wang2022DetectionValidation} a scheme for the protection and isolation of sensor types is explored. In this work, a bank of sensors is used to capture sensor anomalies using a combination of extended Kalman Filters and cumulative summation error monitoring. Comparatively, few works have examined the vulnerabilities of ultrasonic sensing. This is likely due to the lower fidelity offered by ultrasonics compared with LiDAR. In \cite{Lee2019SecuringModel} a solution for securing ultrasonic sensors against signal injection attacks is examined. Various vulnerabilities of ultrasonics were also explored in \cite{Lim2018AutonomousAssessment} wherein Lim \textit{et al.} performed an impact assessment on some ultrasonic attacks using an Arduino-based experimental set-up. 

While these works provide many insights into the vulnerabilities and protection of ToF sensors, they typically offer specialized solutions, for their relative use cases. They also often rely on the integrity of a single measurement. By contrast, our proposed work offers a generalized solution, which uses surrounding measurements to corroborate the measurement integrity. This method can be performed quickly and within the timescales required for collision detection in self-driving cars.


\section{Proposed Model}
\label{sec:formal}
In this section, we outline the formal detection and adversary framework for the proposed \textit{STAnDS} model. We start by outlining the attack models explored in this work. Following this, we outline the 2-stage detection approach used by \emph{STAnDS} for identifying the attacks against ToF distance sensing in AVs. 

\subsection{Attack Model}
We consider $2$ main types of attack in this paper, A) \textit{Triggering attacks},  wherein an object is made to appear closer than it actually is and B) \textit{Deflection style attacks}, where an object is removed from the view. We first describe each of them before going on to discuss the proposed approach \textit{STAnDS}.

\subsubsection{Triggering Attacks} 
\begin{figure}[h]
 \centering
 \includegraphics[width=2.5in]{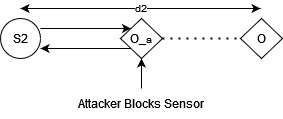}
 \caption{Triggering attack wherein the sensor where a fictitious object is made to appear close to source. }
 \label{triggeringattack}
 \end{figure}

In this attack, the sensor is spuriously triggered to make it look like an object has appeared in front of the sensor. These can be used to intentionally trigger brake sensors erroneously with the intent to cause collisions when travelling at high speed. We illustrate this attack in Figure \ref{triggeringattack} wherein the attacker spoofs the sensor to create a fictitious object $O_a$ to block the signal prior to the actual object being distance measured. As shown in the previous works discussed, \textit{Triggering attacks} can be done using lasers on the LiDAR system. This can be performed with a basic actuator-based bug placed on the AV sensor itself.

\subsubsection{Deflection Attacks} 
For the most part, ToF sensors rely on the response of a pinged signal to confirm the distance of an object. \textit{Triggering attacks} can fool the sensor into thinking an object is closer than it actually is. On the contrary, in \textit{Deflection attacks} the attacker will aim to bounce the return signal away from the receiving sensor or otherwise remove an object from view. As the signal is bounced away, ToF becomes essentially infinite and the operator will believe that there is no object in front of the sensor. We illustrate this type of attack in Figure \ref{deflectionattack} where a curved surface is used to deflect the returning signal upward.

 \begin{figure}[h]
 \centering
 \includegraphics[width=2.5in]{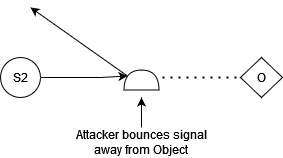}
 \caption{Deflection attack wherein the return signal is reflected away or otherwise hidden from the sensor. }
 \label{deflectionattack}
 \end{figure}

\subsection{STAnDS Detection Approach}
To detect the previously described attacks, we proposed \emph{STAnDS: Spatial-Temporal Anomaly Detection for Sensor}. The proposed model is based on two-stage error checks A) \textit{Residual Error check}, and B) \textit{Temporal Error check} as explained below.

\subsubsection{Residual Error Check}
The initial spatial element is based on a residual error check to ensure the plausibility of the received distance measurement set as they correspond to one another. In this way, the formulation is a generalized approach to sensing protection and can be applied to almost any distance sensing system. 
We consider a static distance sensing system consisting of ToF distance checks. Between a sensor and a distant object in front of the sensor, we consider the actual distance (x-plane) 
$d$, the measured distance $s$ and the implicit error within the measurement $e$ related as given in equation \ref{distance}.

 \begin{equation}
     s = d + e.
     \label{distance}
 \end{equation}

\begin{figure}[h]
 \centering
 \includegraphics[width=2.5in]{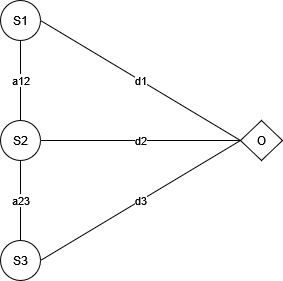}
 \caption{Static sensing setup analysed in this work. }.
 \label{Simplesensoroutline}
 \end{figure}

We consider a system of multiple sensors measuring the same object with the geometric distance between these sensors known and defined by $a$. For initial model simplicity, we consider that the sensors are aligned in a flat plane, checking the object $o$ distance as shown in Figure \ref{Simplesensoroutline}.

We can then establish a co-calculation of the distance using our other available measurements and their relationship such that the correct distance for sensor $S2$ can be corroborated using the distance measurement from sensor $S2$ and the known relationship between the two sensors $S1$ and $S2$.   

 \begin{equation}
     \hat{s_2} =  \sqrt{s_{1}^2-a_{12}^2}
     \label{generalized state equation}
 \end{equation}

Here, $a_{12}$ is the known distance between sensor $S1$ and $S2$. The difference between our estimated sensor measurement and the actual calculated measurement can be used to ensure sensor integrity. For example, we can corroborate sensor $S2$'s distance check using the data for sensor $S1$ such that (the residual error) is computed as given in equation \ref{eq2}.

 \begin{equation}
     r_{1} = s_2 - \hat{s_2}.
     \label{eq2}
 \end{equation}

We can then repeat this for all the relevant sensors and sum the absolutes to establish a cross-system residual as given in equation \ref{eq3}.

 \begin{equation}
     r = \sum{r}_i.
     \label{eq3}
 \end{equation}

\subsubsection{Alarm Limits}
We assume that signal errors $\textbf{x}$  are random, independent and follow a normal distribution with mean zero and unit $\mathcal{N}(0,\sigma^2)$. Therefore, a chi-squared distribution model $\chi^{2}_{m-n,\alpha}$ with $m-n$ degrees of freedom and confidence interval $\alpha$  (typically 0.95 or 0.99) can be used to define the detection threshold as given in equation \ref{eq4}.

\begin{equation}
    \eta = \sigma \sqrt{\chi^{2}_{m-n,\alpha}}. 
    \label{eq4}
\end{equation}

We can then use baseline normal operation modelling of $r$ to establish an alarm value such that if $r_t>\eta$ the alarm will trigger. This kind of residual error monitor is similar to what we might see in a power system state estimation \cite{Monticelli1999StateApproach} to verify sensor integrity.

\subsubsection{Expected Change Monitoring}
In this subsection, we outline stage $2$ of our anomaly detection scheme. This scheme allows for verification of the consecutive measurements to ensure they are plausible from a timing perspective.  While the residual-based approach is good at identifying spatial anomalies between the different sensors, in the past these types of approaches have been shown to be vulnerable to manipulation \cite{Higgins2021TopologyInformation}. Therefore, it is important to verify data on a temporal (or time-based approach) as well as spatial verification. For each respective distance $d$ and time interval $t$ we can calculate the change in distance measured across the time period such that

 \begin{equation}
     \Delta{d}_{12} = d_1-d_2.
     \label{deltad}
 \end{equation}

We can then consider an anomaly based on violation of this expected change such that $\Delta{d}_{12}>E(\Delta{d_{12}})$. Setting these limits is not trivial as the operator needs to balance the risk of not engaging the brakes with the plausibility of an attack or even events which would not warrant a braking intervention (such as a leaf covering the sensor). Event-triggered methodologies such as these can also be augmented using 'Moving Target' style defences such as \cite{Xu2022BlendingApproach} in order to provide hard anomaly checks on top of the data-driven approach.

\section{Results}
\label{sec:evaluation}
This section assesses the performance of spatial-temporal anomaly detection on different attack types. All simulations were performed using an Intel Core i7-7820X CPU with 64GB of RAM running on a Windows 10 system. We simulated a system as given in Figure \ref{Simplesensoroutline} with the system parameters outlined in Table \ref{table1}. We simulate sensor noise or measurement error by adding noise to the actual measured distance such that $z_a =d_a + n_a$ where $n_a$ is a uniform random variable such that $n_a \sim U(0,0.1)$

\begin{table}[]
\begin{tabular}{|lllll|}
\hline
\multicolumn{5}{|c|}{\textbf{Distance Parameters}}                                                                              \\ \hline
\multicolumn{1}{|l|}{a1 (m)} & \multicolumn{1}{l|}{a2 (m)} & \multicolumn{1}{l|}{d1 (m)} & \multicolumn{1}{l|}{d2 (m)} & d3 (m) \\ \hline
\multicolumn{1}{|l|}{1}      & \multicolumn{1}{l|}{2}      & \multicolumn{1}{l|}{10.05}  & \multicolumn{1}{l|}{10}     & 10.2   \\ \hline
\end{tabular}
\label{table1}
\end{table}

\subsection{Normal Operation}

\begin{figure}[h]
 \centering
 \includegraphics[width=2.5in]{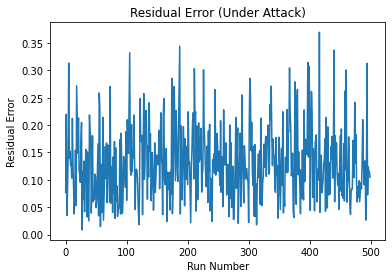}
 \caption{Residual Error Under normal Operation with no attack in place. }.
 \label{residualnormal}
 \end{figure}

\begin{figure}[h]
 \centering
 \includegraphics[width=3.0in]{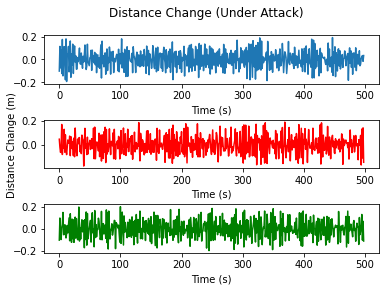}
 \caption{Distance change under normal operation with no attack in place. Different colours represent the different sensors change in distance.  }.
 \label{distancechangenormal}
 \end{figure}

In Figure \ref{residualnormal}, we illustrate the system residual under normal operation. We note the volatile but uniform distribution of the residual over the 500 runs. The average residual level is around $0.13$ with a standard deviation of $0.072$. Under a $2$ standard deviation confidence interval this would give us a residual limit of $0.27$ which would correspond to a false positive detection rate of about $5$\%. We can also see a similar uniformity in Figure \ref{distancechangenormal} which illustrates the distance change under normal operation.

\subsection{Triggering Attack}
\begin{figure}[h]
 \centering
 \includegraphics[width=3.0in]{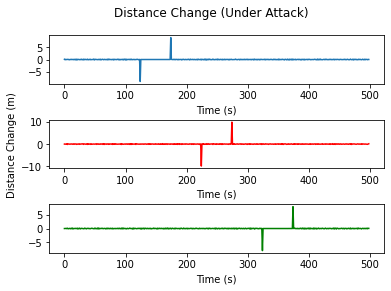}
 \caption{Distance Change under triggering attack. Different colours represent the different sensors change in distance.   }.
 \label{distancechangetrigger}
 \end{figure}

The purpose of the triggering attack is to make the system operator believe a fictitious object is in front of the vehicle by simulating an object in front of the sensor. To replicate what a triggering attack might look like in simulation we overwrite the distance measured to $0.1$ in order to simulate this effect. In Figure \ref{triggeringattack}, we show the residual under attack. We apply three sensor attacks independently. The increase in residual error is clear and should be easily notable for any system operator. We note that the residual error increase is more prevalent for sensor $S2$ than either sensor $S1$ or $S3$. This is because in our model sensors $S1$ \& $S3$ are not independently verified but compared with the expected distance under sensor $S2$. Therefore, when we trigger the sensor we have 2 corroboration points compared with just 1 for either sensor $S1$ or $S3$. Even without the residual error detection the expected change monitoring also provides an indication of an anomalous event. In Figure \ref{distancechangetrigger} we can see the impact of the trigger attacks on the expected system changes. We see larger incident changes as the triggered attacks are committed which can be used to indicate that something is wrong.  

 \begin{figure}[h]
 \centering
 \includegraphics[width=2.5in]{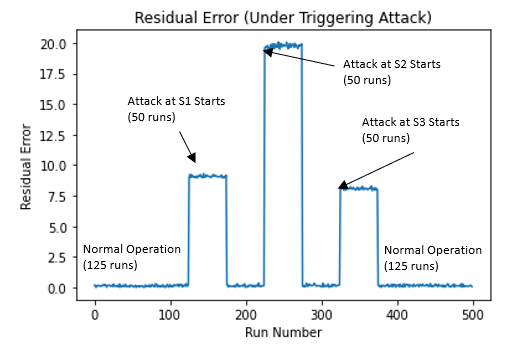}
 \caption{Residual Error under triggering attack. }.
 \label{deflectionattack}
 \end{figure}

\subsection{Deflection Attack}
\begin{figure}[h]
 \centering
 \includegraphics[width=3.0in]{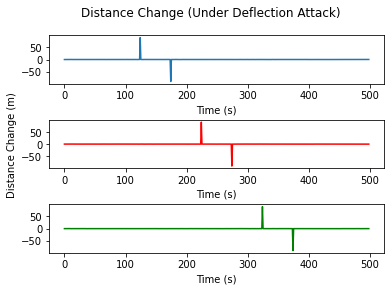}
 \caption{Distance Change under deflection attack. Different colours represent the different sensors change in distance.  }.
 \label{distancechangedeflection}
 \end{figure}

For the deflection attack, we overwrite the sensor with a large distance (in this case $100m$) to simulate a signal being bounced away from the receiver or otherwise hidden. We see a similar result pattern with the deflection attack results as we did with the triggering attack. This is shown in Figure \ref{deflectionattack}, where we can see the elevated residual error under the attack occurring at run $150$. In Figure \ref{distancechangedeflection}, we can see the large distance change increase as a result of the deflection attack which would likely result in an anomaly being raised to the system operator. We note that the acceptable spatial relationships will be much more robustly defined based on the system noise. The temporal distance checking would benefit from a practical analysis which can distinguish between genuine breaking changes and those introduced by spoofing.

 \begin{figure}[h]
 \centering
 \includegraphics[width=2.5in]{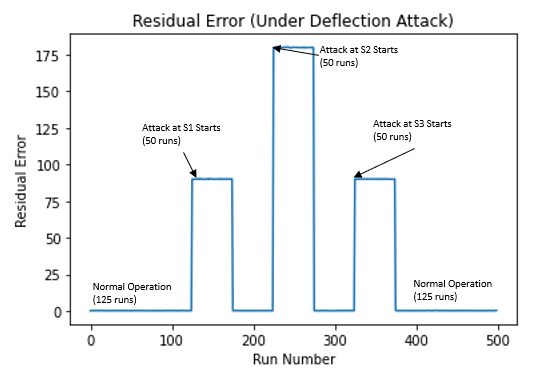}
 \caption{Residual Error under deflection attack. }.
 \label{deflectionattack}
 \end{figure}

\section{Conclusions \& Future Work}
\label{sec:conclusion}
In this work, we examined how ToF sensing systems can be enhanced to be resilient to deflection and trigger-style attacks using a spatial-temporal anomaly detection approach. The approach protects sensors from attacks using spatial-temporal layered anomaly detection. This approach has several advantages over the current state-of-the-art including low latency \& the ability to be generalized across many sensing types and different systems. In future work, we intend to test these checks on a practical sensing test-bed to confirm their efficacy and reliability in real-life scenarios.

\section*{Acknowledgment}
This work was supported by UK Research and Innovation’s Digital Security by Design (DSbD) project.

\ifCLASSOPTIONcaptionsoff
  \newpage
\fi

\bibliographystyle{IEEEtran.bst}

\bibliography{bare_jrnl.bbl}

\begin{thebibliography}{10}
\providecommand{\url}[1]{#1}
\csname url@samestyle\endcsname
\providecommand{\newblock}{\relax}
\providecommand{\bibinfo}[2]{#2}
\providecommand{\BIBentrySTDinterwordspacing}{\spaceskip=0pt\relax}
\providecommand{\BIBentryALTinterwordstretchfactor}{4}
\providecommand{\BIBentryALTinterwordspacing}{\spaceskip=\fontdimen2\font plus
\BIBentryALTinterwordstretchfactor\fontdimen3\font minus
  \fontdimen4\font\relax}
\providecommand{\BIBforeignlanguage}[2]{{%
\expandafter\ifx\csname l@#1\endcsname\relax
\typeout{** WARNING: IEEEtran.bst: No hyphenation pattern has been}%
\typeout{** loaded for the language `#1'. Using the pattern for}%
\typeout{** the default language instead.}%
\else
\language=\csname l@#1\endcsname
\fi
#2}}
\providecommand{\BIBdecl}{\relax}
\BIBdecl

\bibitem{Thing2017AutonomousDefences}
V.~L. Thing and J.~Wu, ``{Autonomous Vehicle Security: A Taxonomy of Attacks
  and Defences},'' in \emph{Proceedings - 2016 IEEE International Conference on
  Internet of Things; IEEE Green Computing and Communications; IEEE Cyber,
  Physical, and Social Computing; IEEE Smart Data,
  iThings-GreenCom-CPSCom-Smart Data 2016}.\hskip 1em plus 0.5em minus
  0.4em\relax Institute of Electrical and Electronics Engineers Inc., 5 2017,
  pp. 164--170.

\bibitem{Sun2022ACAVs}
X.~Sun, F.~R. Yu, and P.~Zhang, ``{A Survey on Cyber-Security of Connected and
  Autonomous Vehicles (CAVs)},'' \emph{IEEE Transactions on Intelligent
  Transportation Systems}, vol.~23, no.~7, pp. 6240--6259, 7 2022.

\bibitem{Kim2021CybersecurityDefense}
K.~Kim, J.~S. Kim, S.~Jeong, J.~H. Park, and H.~K. Kim, ``{Cybersecurity for
  autonomous vehicles: Review of attacks and defense},'' 4 2021.

\bibitem{Kohli2018EnablingCrash}
\BIBentryALTinterwordspacing
P.~Kohli and A.~Chadha, ``{Enabling Pedestrian Safety using Computer Vision
  Techniques: A Case Study of the 2018 Uber Inc. Self-driving Car Crash},'' 5
  2018. [Online]. Available: \url{http://arxiv.org/abs/1805.11815
  http://dx.doi.org/10.1007/978-3-030-12388-8_19}
\BIBentrySTDinterwordspacing

\bibitem{Hong2021AICars}
J.~W. Hong, I.~Cruz, and D.~Williams, ``{AI, you can drive my car: How we
  evaluate human drivers vs. self-driving cars},'' \emph{Computers in Human
  Behavior}, vol. 125, 12 2021.

\bibitem{PetitRemoteLiDAR}
J.~Petit, B.~Stottelaar, M.~Feiri, and F.~Kargl, ``{Remote Attacks on Automated
  Vehicles Sensors: Experiments on Camera and LiDAR},'' Tech. Rep.

\bibitem{Pham2021AVehicles}
M.~Pham and K.~Xiong, ``{A survey on security attacks and defense techniques
  for connected and autonomous vehicles},'' 10 2021.

\bibitem{Li2021FoolingPerturbation}
Y.~Li, C.~Wen, F.~Juefei-Xu, and C.~Feng, ``{Fooling LiDAR Perception via
  Adversarial Trajectory Perturbation},'' in \emph{Proceedings of the IEEE
  International Conference on Computer Vision}.\hskip 1em plus 0.5em minus
  0.4em\relax Institute of Electrical and Electronics Engineers Inc., 2021, pp.
  7878--7887.

\bibitem{Cao2019AdversarialDriving}
Y.~Cao, Y.~Zhou, Q.~A. Chen, C.~Xiao, W.~Park, K.~Fu, B.~Cyr, S.~Rampazzi, and
  Z.~Morley~Mao, ``{Adversarial sensor attack on LiDAR-based perception in
  autonomous driving},'' in \emph{Proceedings of the ACM Conference on Computer
  and Communications Security}.\hskip 1em plus 0.5em minus 0.4em\relax
  Association for Computing Machinery, 11 2019, pp. 2267--2281.

\bibitem{Kamal2021GPSVehicles}
M.~Kamal, A.~Barua, C.~Vitale, C.~Laoudias, and G.~Ellinas, ``{GPS Location
  Spoofing Attack Detection for Enhancing the Security of Autonomous
  Vehicles},'' in \emph{IEEE Vehicular Technology Conference}, vol.
  2021-September.\hskip 1em plus 0.5em minus 0.4em\relax Institute of
  Electrical and Electronics Engineers Inc., 2021.

\bibitem{Hallyburton2021SecurityVehicles}
\BIBentryALTinterwordspacing
R.~S. Hallyburton, Y.~Liu, Y.~Cao, Z.~M. Mao, and M.~Pajic, ``{Security
  Analysis of Camera-LiDAR Fusion Against Black-Box Attacks on Autonomous
  Vehicles},'' 6 2021. [Online]. Available:
  \url{http://arxiv.org/abs/2106.07098}
\BIBentrySTDinterwordspacing

\bibitem{Alheeti2022LiDARVehicles}
K.~M. Alheeti, A.~Alzahrani, and D.~Al~Dosary, ``{LiDAR Spoofing Attack
  Detection in Autonomous Vehicles},'' in \emph{Digest of Technical Papers -
  IEEE International Conference on Consumer Electronics}, vol.
  2022-January.\hskip 1em plus 0.5em minus 0.4em\relax Institute of Electrical
  and Electronics Engineers Inc., 2022.

\bibitem{Sun2020TowardsCountermeasures}
\BIBentryALTinterwordspacing
J.~Sun, Y.~Cao, Q.~A. Chen, and Z.~M. Mao, ``{Towards Robust LiDAR-based
  Perception in Autonomous Driving: General Black-box Adversarial Sensor Attack
  and Countermeasures Towards Robust LiDAR-based Perception in Autonomous
  Driving: General Black-box Adversarial Sensor Attack and Countermeasures},''
  in \emph{USENIX Security Symposium}, 2020. [Online]. Available:
  \url{https://www.usenix.org/conference/usenixsecurity20/presentation/sun}
\BIBentrySTDinterwordspacing

\bibitem{Wang2022DetectionValidation}
Y.~Wang, Q.~Liu, E.~Mihankhah, C.~Lv, and D.~Wang, ``{Detection and Isolation
  of Sensor Attacks for Autonomous Vehicles: Framework, Algorithms, and
  Validation},'' \emph{IEEE Transactions on Intelligent Transportation
  Systems}, vol.~23, no.~7, pp. 8247--8259, 7 2022.

\bibitem{Shafique2021DetectingModels}
A.~Shafique, A.~Mehmood, and M.~Elhadef, ``{Detecting Signal Spoofing Attack in
  UAVs Using Machine Learning Models},'' \emph{IEEE Access}, vol.~9, pp.
  93\,803--93\,815, 2021.

\bibitem{Cao2021Demo:Targets}
Y.~Cao, J.~Ma, K.~Fu, S.~Rampazzi, and M.~Mao, ``{Demo: Automated Tracking
  System For LiDAR Spoofing Attacks On Moving Targets}.''\hskip 1em plus 0.5em
  minus 0.4em\relax Internet Society, 8 2021.

\bibitem{Lee2019SecuringModel}
\BIBentryALTinterwordspacing
S.~Lee, W.~Choi, and D.~H. Lee, ``{Securing Ultrasonic Sensors Against Signal
  Injection Attacks Based on a Mathematical Model},'' \emph{IEEE Access},
  vol.~7, pp. 107\,716--107\,729, 2019. [Online]. Available:
  \url{https://ieeexplore.ieee.org/document/8786810/}
\BIBentrySTDinterwordspacing

\bibitem{Lim2018AutonomousAssessment}
B.~S. Lim, S.~L. Keoh, and V.~L. Thing, ``{Autonomous vehicle ultrasonic sensor
  vulnerability and impact assessment},'' in \emph{IEEE World Forum on Internet
  of Things, WF-IoT 2018 - Proceedings}, vol. 2018-January.\hskip 1em plus
  0.5em minus 0.4em\relax Institute of Electrical and Electronics Engineers
  Inc., 5 2018, pp. 231--236.

\bibitem{Monticelli1999StateApproach}
A.~Monticelli, \emph{{State Estimation in Electric Power Systems: A Generalized
  approach}}, 1st~ed.\hskip 1em plus 0.5em minus 0.4em\relax Springer, 5 1999,
  vol.~1.

\bibitem{Higgins2021TopologyInformation}
\BIBentryALTinterwordspacing
M.~Higgins, J.~Zhang, N.~Zhang, and F.~Teng, ``{Topology Learning Aided False
  Data Injection Attack without Prior Topology Information},'' in \emph{IEEE
  PES General Meeting (GM)}, 7 2021, pp. 1--1. [Online]. Available:
  \url{http://arxiv.org/abs/2102.12248}
\BIBentrySTDinterwordspacing

\bibitem{Xu2022BlendingApproach}
\BIBentryALTinterwordspacing
W.~Xu, M.~Higgins, J.~Wang, I.~M. Jaimoukha, and F.~Teng, ``{Blending Data and
  Physics Against False Data Injection Attack: An Event-Triggered Moving Target
  Defence Approach},'' 4 2022. [Online]. Available:
  \url{http://arxiv.org/abs/2204.12970}
\BIBentrySTDinterwordspacing

\end{thebibliography}




\end{document}